\documentclass[12pt,onecolumn]{IEEEtran} 

\usepackage{amsmath,amsopn,amssymb}
\usepackage{graphicx,xspace,color}
\usepackage{epsfig,subfigure}
\usepackage{longtable,multirow}
\usepackage{array,float}
\usepackage{cite}
\usepackage[nolist]{acronym}
\usepackage{soul}
\usepackage{booktabs}
\usepackage[font=small]{caption}

\newtheorem{lemma}{Lemma}
\graphicspath{ {./Figures/}}

\begin{document}

\begin{acronym}
\acro{IA}{interference alignment}
\acro{CSI}{channel state information}
\acro{SNR}{signal-to-noise ratio}
\acro{MIMO}{multiple-input multiple-output}
\acro{DoF}{degrees of freedom}
\acro{iid}{independent and identically distributed}
\acro{TDD}{time division duplexing}
\acro{TDMA}{Time Division Multiple Access}
\acro{SU-MIMO}{Single-User MIMO}
\acro{ZF}{zero-forcing}
\acro{SINR}{signal-to-noise-plus-interference ratio}
\acro{MMSE}{minimum mean square error}
\acro{MI}{mutual information}
\end{acronym}

\title{Breaking the Interference Barrier in Dense Wireless Networks with Interference Alignment}
\author{Konstantinos Dovelos and Boris Bellalta\\
\fontsize{10}{10}
Universitat Pompeu Fabra (UPF), Barcelona}

\maketitle
\thispagestyle{empty}

\baselineskip20pt

\begin{abstract}
\baselineskip20pt

A fundamental problem arising in dense wireless networks is the high co-channel interference. Interference alignment (IA) was recently proposed as an effective way to combat interference in wireless networks. The concept of IA, though, is originated by the capacity study of interference channels and as such, its performance is mainly gauged under ideal assumptions, such as instantaneous and perfect \ac{CSI} at all nodes, and a homogeneous \ac{SNR} regime, i.e., each user has the same average \ac{SNR}. Consequently, the performance of IA under realistic conditions has not been completely investigated yet. In this paper, we aim at filling this gap by providing a performance assessment of spatial IA in practical systems. Specifically, we derive a closed-form expression for the IA average sum-rate when \ac{CSI} is acquired through training and users have heterogeneous \ac{SNR}. A main insight from our analysis is that IA can indeed provide significant spectral efficiency gains over traditional approaches in a wide range of dense network scenarios. To demonstrate this, we consider the examples of linear, grid and random network topologies.
\end{abstract}

\section{Introduction}
The extreme traffic load future wireless networks are expected to accommodate requires a re-thinking of the system design. A traditional approach to deal with interference from coexisting networks, i.e., networks operating at the same frequency band, is to avoid it or minimize it through sophisticated spectrum allocation and power control. Keeping co-channel interference sufficiently low, enables users to reliably decode their signals by treating interference as noise \cite{Low_IF_Regime}. However, as the network infrastructure becomes more dense to fulfill the demand for fast and ubiquitous wireless connectivity, co-channel interference becomes severe and limits the system performance \cite{IFL_Systems}. As a result, interference management will play a key role in the design of future wireless systems. 

Towards this end, \ac{IA} was recently proposed as an effective technique to deal with co-channel interference. The main idea behind spatial \ac{IA} is to exploit the additional signaling dimensions provided by \ac{MIMO} to mitigate interference. Specifically, a fraction of the antennas at each receiver is used to confine the interference into a lower dimensional signal space, rendering the rest of the signaling dimensions available for interference-free communication. Early theoretical studies have shown that \ac{IA} is the optimal (in terms of spatial multiplexing gain) transmission/receive strategy for the interference channel at the high \acf{SNR} regime \cite{IA_KUser_IFC}. However, the performance of \ac{IA} is mostly gauged in terms of the multiplexing gain, which is an asymptotic performance metric and as such, it does not capture large-scale fading that might yield heterogeneous \ac{SNR} across the users, i.e., heterogeneous \ac{SNR} regime. Furthermore, instantaneous and global channel knowledge at all nodes is assumed, which is not true in practical systems where \acf{CSI} is acquired through training and comes with additional overheads, yielding further performance losses. There are a few works in the literature suggesting that the performance of \ac{IA} might not be as good as expected in the heterogeneous \ac{SNR} regime, with either perfect or imperfect \ac{CSI} \cite{SL_PA_IA,IA_RandomNets, IA_CC_ICSI}. This motivate us to further study the performance of \ac{IA} under realistic conditions, and especially in dense network scenarios where the potential benefits of \ac{IA} can be fully exploited. In our performance analysis, we consider the following:
\begin{itemize}
\item Large-scale fading
\item Imperfect \ac{CSI} and training overhead
\item Practical \ac{SNR} values
\end{itemize}
To compare the performance of \ac{IA} against the state-of-the-art, we also consider the following schemes:
\begin{itemize}
\item \ac{SU-MIMO}: All the antennas at each node are used for communication, while multi-user interference is treated as noise.
\item \ac{TDMA}: Transmitters are coordinating their transmissions in time. Each user gets a fraction of the time to transmit and interference is avoided.
\end{itemize}

We begin by deriving a closed-form expression for the \ac{IA} average sum-rate in genie-aided systems with perfect \ac{CSI}, and extend the analysis under a general model for imperfect \ac{CSI}. We then specialize our results to \ac{TDD} systems, where channel reciprocity holds. In this case, global \ac{CSI} can be obtained at each node through training of the reverse and forward channels, eliminating the need for feedback. Next, we compute the achievable user rates for \ac{IA} under \ac{CSI} mismatch, when receivers ignore channel estimation errors (assumption 1) and employ nearest neighbor decoding (assumption 2). Under assumptions 1 and 2, it has been proven that regardless of the interference distribution, a lower-bound on the achievable rate can be derived, which corresponds to the minimum achievable rate of a typical receiver \cite{NND_NonGC,MIMO_Training, MIMO_Capacity_ICSI}. Using this fact, we finally derive a closed-form expression for the average effective sum-rate of \ac{IA}, \ac{TDMA} and \ac{SU-MIMO} and run computer simulations for a variety of network topologies.

Throughout this paper, we use the following notation: $\mathbf{A}$ is a matrix; $\mathbf{a}$ is a vector; a is a scalar; $(\cdot)^*$ denotes the conjugate transpose; $\mathbf{I}_N$ is the $N\times N$ identity matrix; $\mathcal{CN}(\mu, \sigma^2)$ is a complex Gaussian random variable with mean $\mu$ and variance $\sigma^2$; $\mathbb{E[\cdot]}$ denotes expectation.

\begin{figure}[H]
	\centering
	\includegraphics[scale=1]{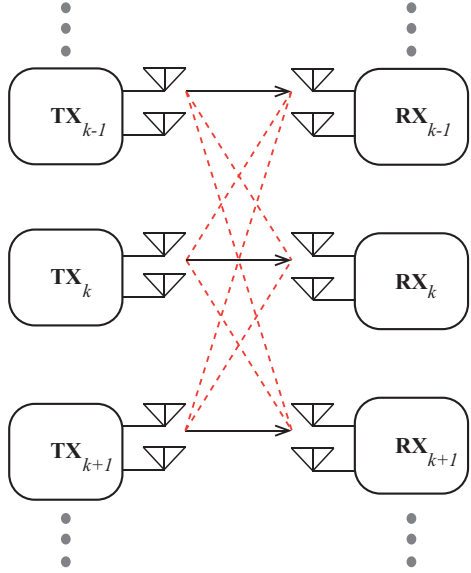}
	\caption{K-user \ac{MIMO} interference channel model. Solid black arrows and dash red lines represent the direct channels and the interfering channels, respectively.}
	\label{fig:3UserIFC}
\end{figure}

\section{System Model}
Consider a \ac{MIMO} interference network consisting of $K$ transmitter-receiver pairs, as shown in Fig. \ref{fig:3UserIFC}. All transmitters and receivers are equipped with $N_t$ and $N_r$ antennas, respectively. Each transmitter seeks to communicate $d \leq \min(N_r, N_t)$ spatial streams to its paired receiver. The wireless channels are characterized by small-scale and large-scale fading. Let $\eta_{ki}$ represent the large-scale fading of the link $(k,i)$. Furthermore, let $\mathbf{H}_{ki}\in \mathbb{C}^{N_r\times N_t}$ be the \ac{MIMO} Rayleigh-fading channel with \ac{iid} complex Gaussian entries of zero mean and unit variance. The channel from transmitter $i$ to receiver $k$ is then modeled as $\sqrt{\eta_{ki}}\ \mathbf{H}_{ki}$. Assuming perfect synchronization, the sampled baseband signal at receiver $k$ is given by
\begin{align*}
\mathbf{y}_k &= \sum_{i=1}^{K}\sqrt{\frac{P \eta_{ki}}{d}}\mathbf{H}_{ki}\mathbf{V}_{i}\mathbf{s}_i +\mathbf{n}_k \\
& = \sqrt{\frac{P \eta_{kk}}{d}}\mathbf{H}_{kk}\mathbf{V}_k\mathbf{s}_k +  \underbrace{\sum_{i=1,i\neq k}^{K}\sqrt{\frac{P \eta_{ki}}{d}}\mathbf{H}_{ki}\mathbf{V}_i\mathbf{s}_i}_{\text{interference}} +\mathbf{n}_k
\end{align*}
where $\mathbf{V}_k\in \mathbb{C}^{N_t\times d}$ and $\mathbf{s}_k \in \mathbb{C}^{d\times 1}$ are the precoding matrix and the data symbol vector of transmitter $k$, respectively, $P$ is the transmit power per channel access and $\mathbf{n}_k$ is the complex white Gaussian noise with covariance matrix $\mathbb{E}[\mathbf{n}_k\mathbf{n}_k^*] = \sigma_0^2\mathbf{I}_{N_r}$. The data symbol vector of transmitter $k$ is drawn from a Gaussian codebook and has \ac{iid} entries of zero mean and variance one, i.e., $\mathbf{E}[\mathbf{s}_k\mathbf{s}_k^*] = \mathbf{I}_d$.

\section{Interference Alignment}
\subsection{Perfect \ac{CSI}}
In \ac{IA}, receiver $k$ obtains an estimate of $\mathbf{s}_k$ as
\[
\hat{\mathbf{s}}_k = \mathbf{U}^*_k\mathbf{y}_k
\]
where $\mathbf{U}_k\in\mathbb{C}^{N_r\times d}$ is the combining matrix. Assuming perfect \ac{CSI}, the precoders $\{\mathbf{V}_k\}_{k=1}^K$ and combiners $\{\mathbf{U}_k\}_{k=1}^K$  are jointly designed to be \textit{unitary} and to satisfy the following conditions:
\begin{align*}\label{ia_conditions}
\mathbf{U}^*_k \mathbf{H}_{ki}\mathbf{V}_i &= \mathbf{0},\quad \forall i\neq k  \\
\text{rank} \left(\mathbf{U}^*_k\mathbf{H}_{kk} \mathbf{V}_k\right) &= d
\end{align*}
If so, the interfering signals at each receiver are aligned to a subspace of the received signal space and can be suppressed by zero-forcing (ZF) combining\footnote{An iterative algorithm for designing the precoders and \ac{ZF} combiners is the minimum interference leakage algorithm in \cite{DIA}.}, as shown in Fig. 2. Specifically, the post-processed signal at receiver k is written as

\begin{align*}
\hat{\mathbf{s}}_k &= \sqrt{\frac{P \eta_{kk}}{d}}\mathbf{U}^*_k\mathbf{H}_{kk}\mathbf{V}_k\mathbf{s}_k + \sum_{i\neq k}^{K}\sqrt{\frac{P \eta_{ki}}{d}}\mathbf{U}^*_k\mathbf{H}_{ki}\mathbf{V}_i\mathbf{s}_i+ \mathbf{U}^*_k\mathbf{n}_k\\
&= \sqrt{\frac{P \eta_{kk}}{d}}\mathbf{U}^*_k\mathbf{H}_{kk}\mathbf{V}_k\mathbf{s}_k  + \tilde{\mathbf{n}}_k \\
&= \sqrt{\frac{P \eta_{kk}}{d}}\tilde{\mathbf{H}}_{kk}\mathbf{s}_k + \tilde{\mathbf{n}}_k
\end{align*}
and the desired signal $\mathbf{s}_k$ is received in the absence of interference. The effective $d\times d$ channel $\tilde{\mathbf{H}}_{kk} = \mathbf{U}_k^*\mathbf{H}_{kk}\mathbf{V}_k$ and the effective noise $\tilde{\mathbf{n}}_k = \mathbf{U}_k^*\mathbf{n}_k$ have \ac{iid} entries  following $\mathcal{CN}(0,1)$ and $\mathcal{CN}(0,\sigma_0^2)$, respectively. This holds because matrices $\mathbf{U}_k$ and $\mathbf{V}_k$ are unitary and independent of both $\mathbf{H}_k$ and $\mathbf{n}_k$, i.e., unitarily invariant matrix/vector \cite{RM_Theory}. As a result, this is a \ac{MIMO} Rayleigh-fading channel and the average achievable rate of user $k$, under joint decoding of the $d$ streams, is given by \cite{CF_Capacity}

\begin{align*}
\bar{R}_k  =  \mathbb{E}\left[\log \text{det}\left(\mathbf{I}_{d} + \frac{P \eta_{kk}}{\sigma_0^2d}\tilde{\mathbf{H}}_{kk}\tilde{\mathbf{H}}_{kk}^*
\right) \right] = f\left(\frac{P \eta_{kk}}{\sigma_0^2}, d\right)
\end{align*}

where 

\begin{align*}
f(x, y) = \log_2(e)e^{y/x}\sum_{i=0}^{d-1}\sum_{j=0}^{i}\sum_{l=0}^{2j}\frac{(-1)^l (2j)!(m-d+l)!}{2^{2i-l}j! l!(m-d+j)!} \binom{2i-2j}{i-j}\binom{2j+2m-2d}{2j-l}\sum_{n=0}^{m-d+l}\text{E}_{n+1}\left(\frac{y}{x}\right)
\end{align*}
\\
with $d=\min(N_r,N_t)$, $m=\max(N_r,N_t)$ and $E_{n}(z) = \int_{1}^{\infty}e^{-zx}x^{-n}dx$ being the exponential integral of order $n$. Finally, the average sum rate attained over the interference channel is the sum of the individual rates, i.e., $\sum_{k=1}^{K}\bar{R}_k$. 
\\
\begin{figure}[H]
	\centering
	\includegraphics[scale=1.4]{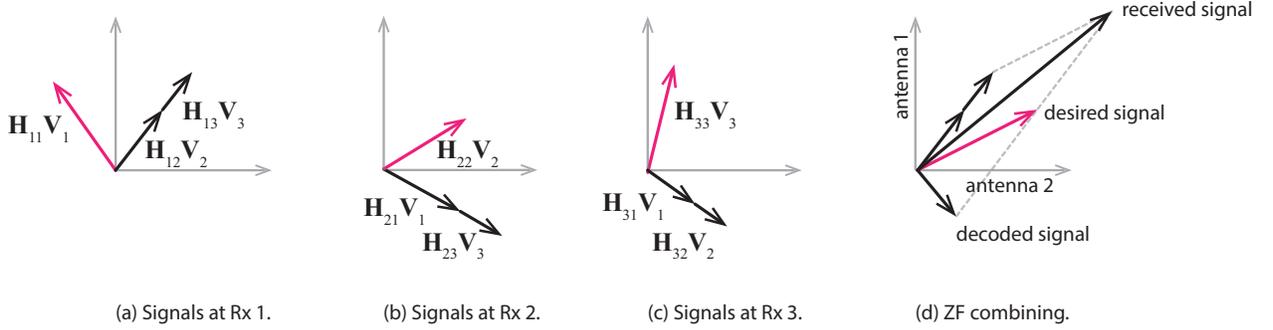}
	\caption{Schematic representation of \ac{IA} for the 3-user $2\times 2$ \ac{MIMO} interference channel under perfect \ac{CSI}.}
	\label{fig:IA_Perfect_CSI}
\end{figure}

\subsection{Imperfect CSI and Training Overhead}\label{subsec:perf_analysis}
With precoders and receive filters obtained on the basis of noisy channel estimates, signals become misaligned, yielding interference at the receivers and a respective loss in the achievable rate. Additionally, the overhead of pilot transmissions might significantly degrade the maximum achievable rate of IA, especially when the coherence interval of the channel is short, i.e, fast fading. In this subsection, we examine the effect of imperfect \ac{CSI} and the associated overheads on the performance of \ac{IA}.

We adopt a block-fading model wherein the channels remain fixed for a period of time denoted by $\tau_{\text{coh}}$, but vary independently from block to block. Furthermore, we focus on \ac{TDD} where channel reciprocity holds. This implies that the forward channel $\mathbf{H}_{ki}$ is equal to $\mathbf{H}^{\text{r}*}_{ik}$, where $\mathbf{H}^{\text{r}*}_{ik}$ denotes the reverse channel from receiver $k$ to transmitter $i$. The coherence interval is divided into the following three phases:
\begin{enumerate}
\item \textbf{Reverse training}: Each receiver $k$ broadcasts an orthogonal pilot sequence matrix $\mathbf{\Phi}_k\in\mathbb{C}^{N_r\times \tau_{\text{rp}}}$ such that $\mathbf{\Phi}_k\mathbf{\Phi}_m^* = \delta_{kj}\mathbf{I}_{N_r}$, spanning $\tau_{\text{rp}} \geq KN_r$ symbols. Each transmitter $i$ then estimates the reverse channels $\hat{\mathbf{H}}^{\text{r}*}_{ik}$ and infer the forward channels as 
\[
\hat{\mathbf{H}}_{ki}= \hat{\mathbf{H}}^{\text{r}*}_{ik}, \quad k=1,\dots, K.
\] 
After the training phase, transmitters report their \ac{CSI} to a central unit, where the precoders $\hat{\mathbf{V}}_k$ and the combiners $\hat{\mathbf{W}}_k$ are computed on the basis of the channel estimates such as the \ac{IA} conditions are fulfilled, 
\begin{align}\label{ia_conditions_ap}
\hat{\mathbf{W}}^*_{k} \hat{\mathbf{H}}_{ki}\hat{\mathbf{V}}_{i} &= \mathbf{0},\quad \forall i\neq k\\\nonumber
\text{rank}\left(\hat{\mathbf{W}}^*_{k} \hat{\mathbf{H}}_{kk}\hat{\mathbf{V}}_{k}\right) &=d
\end{align}
\item \textbf{Forward training}: Each transmitter $i$ transmits an orthogonal pilot matrix $\mathbf{\Omega}_i\in\mathbb{C}^{d\times \tau_{\text{fp}}}$ such that $\mathbf{\Omega}_i\mathbf{\Omega}_j^* = \delta_{ij}\mathbf{I}_{d}$, spanning $\tau_{\text{fp}}\geq Kd$ symbols. These symbols are precoded to enable the estimation of the precoded channels $\mathbf{A}_{ki} = \mathbf{H}_{ki}\hat{\mathbf{V}}_i, i=1,\dots,K$, at each user $k$. Specifically, the observed signal at receiver $k$ is 
\[
\mathbf{Y}_k =  \sum_{i=1}^K\sqrt{\frac{\tau_{\text{fp}} P\eta_{ki}}{d}} \mathbf{H}_{ki}\hat{\mathbf{V}}_i\mathbf{\Omega}_i + \mathbf{Z}_k
\]
where $\mathbf{Z}_k\in\mathbb{C}^{N_r \times\tau_{\text{fp}}}$ is the matrix of noise terms. Since $\hat{\mathbf{V}}_i$ is unitary and independent of the $\mathbf{H}_{ki}$, it follows that $\mathbf{A}_{ki}$ is complex Gaussian. The \ac{MMSE} estimate of $\mathbf{A}_{ki}$ is given by \cite{FT_CSI}
\[
\hat{\mathbf{A}}_{ki} = \frac{\sqrt{\frac{\tau_{\text{fp}} P\eta_{ki}}{d}}}{\sigma_0^2 + \frac{\tau_{\text{fp}} P\eta_{ki}}{d}}\mathbf{Y}_k\mathbf{\Omega}_i^*
\]
and the corresponding estimation error $\mathcal{E}_{ki} = \mathbf{A}_{ki} - \hat{\mathbf{A}}_{ki}$ is zero-mean complex Gaussian with variance
\[
\sigma_{\mathcal{E}_{ki}}^2 = \frac{\sigma_0^2}{\sigma_0^2 +\frac{\tau_{\text{fp}} P\eta_{ki}}{d}} 
\]

whereas $\hat{\mathbf{A}}_{ki}$ has entries with variance $1-\sigma_{\mathcal{E}_{ki}}^2$.  After the forward training stage, each receiver $k$ computes its receive filter $\hat{\mathbf{U}}_k$ on the basis of the precoded channel estimates such that
\begin{align}\label{ia_conditions_u}
\hat{\mathbf{U}}^*_{k} \hat{\mathbf{A}}_{ki} &= \mathbf{0},\quad \forall i\neq k\\\nonumber
\text{rank}\left(\hat{\mathbf{U}}^*_{k} \hat{\mathbf{A}}_{kk}\right) &=d
\end{align}
\item \textbf{Data transmission}: Transmitters send their data symbol vectors using the precoders computed in \eqref{ia_conditions_ap} and receivers decode the received signal using the combiners computed in \eqref{ia_conditions_u}. Specifically, we have

\begin{align*}\label{eq:signal_csi_error}
\hat{\mathbf{s}}_k &= \sum_{i=1}^K\sqrt{\frac{P\eta_{ki}}{d}} \hat{\mathbf{U}}_{k}^* \mathbf{H}_{ki}\hat{\mathbf{V}}_i \mathbf{s}_i + \hat{\mathbf{U}}_{k}^*\mathbf{n}_k \nonumber \\
&=\sum_{i=1}^K\sqrt{\frac{P\eta_{ki}}{d}} \hat{\mathbf{U}}_{k}^* \mathbf{A}_{ki} \mathbf{s}_i + \hat{\mathbf{U}}_{k}^*\mathbf{n}_k\nonumber  \\
&=\sum_{i=1}^K\sqrt{\frac{P\eta_{ki}}{d}} \hat{\mathbf{U}}_{k}^* \left(\hat{\mathbf{A}}_{ki} + \mathcal{E}_{ki}\right) \mathbf{s}_i + \hat{\mathbf{U}}_{k}^*\mathbf{n}_k  \nonumber \\
&=\sqrt{\frac{P\eta_{kk}}{d}} \hat{\mathbf{U}}_{k}^* \hat{\mathbf{A}}_{kk}\mathbf{s}_k + \underbrace{\sum_{i=1}^K\sqrt{\frac{P\eta_{ki}}{d}} \hat{\mathbf{U}}_{k}^* \mathcal{E}_{ki}\mathbf{s}_i}_{\text{CSI interference}} + \hat{\mathbf{U}}_{k}^*\mathbf{n}_k  \nonumber \\
&=\sqrt{\frac{P\eta_{kk}(1-\sigma^2_{\mathcal{E}_{kk}})}{d}} \tilde{\mathbf{H}}_{kk}\mathbf{s}_k + \tilde{\mathbf{n}}_k 
\end{align*}
\\
where $\tilde{\mathbf{H}}_{kk} = 1/ \sqrt{(1-\sigma_{\mathcal{E}_{kk}}^2)}\hat{\mathbf{U}}^*_k\hat{\mathbf{A}}_{kk}$ is the normalized precoded channel estimate with entries following $\mathcal{CN}(0,1)$ and $\tilde{\mathbf{n}}_k = \sum_{i=1}^K\sqrt{\frac{P\eta_{ki}}{d}} \hat{\mathbf{U}}_{k}^* \mathcal{E}_{ki}\mathbf{s}_i + \hat{\mathbf{U}}_{k}^*\mathbf{n}_k$ is the effective noise. The entries of the matrices $\mathbf{U}_{k}^* \mathcal{E}_{ki}$ are \ac{iid} Gaussian with zero mean and variance $\sigma^2_{\mathcal{E}_{ki}}$. The covariance matrix of the effective noise $\tilde{\mathbf{n}}_k$ is then given by 
\\
\begin{equation*}
\mathbf{Q}_k = \mathbb{E}\left[\tilde{\mathbf{n}}_k\tilde{\mathbf{n}}_k^* \right] = \left(\sum_{i=1}^KP\eta_{ki}\sigma_{\mathcal{E}_{ki}}^2 + \sigma_0^2\right)\mathbf{I}_{d}.
\end{equation*}
\\
The average achievable rate of user $k$, when the receiver ignores channel estimation errors and performs nearest neighbor decoding, is given by \cite{NND_NonGC,MIMO_Training}
\begin{align*}
\bar{R}_k & = \mathbb{E}\left[\log\text{det}\left(\mathbf{I}_{d} +\frac{P\eta_{kk}(1-\sigma_{\mathcal{E}_{kk}}^2)}{d}\mathbf{Q}_k^{-1}\tilde{\mathbf{H}}_{kk} \tilde{\mathbf{H}}_{kk}^*\right)\right] \\\\
&= \mathbb{E}\left[\log\text{det}\left(\mathbf{I}_{d} +\frac{\textsf{snr}_k}{d}\tilde{\mathbf{H}}_{kk} \tilde{\mathbf{H}}_{kk}^*\right)\right]\\\\
& =f\left(\textsf{snr}_k, d\right) 
\end{align*}
\\
where
\[
\textsf{snr}_k = \frac{P\eta_{kk}(1-\sigma_{\mathcal{E}_{kk}}^2)}{\sum_{i=1}^KP\eta_{ki}\sigma_{\mathcal{E}_{ki}}^2 + \sigma_0^2}
\]
is the effective \ac{SNR}. Finally, taking into account a minimum training overhead, i.e., $\tau_{\text{rp}} = KN_r$ and $\tau_{\text{fp}} = Kd$, we derive the effective average sum-rate, 

\begin{equation}\label{eq:se_ia}
\bar{R}^{\textsc{ia}}_{\text{sum}} = \frac{\tau_{\text{coh}} - K(N_r + d)}{\tau_{\text{coh}}}\ \sum_{k=1}^K  f\left(\textsf{snr}_k, d\right).
\end{equation}
~
\end{enumerate}

\section{Results}
We quantitatively study the performance of \ac{IA} and compare it to that of \ac{TDMA} and \ac{SU-MIMO}. We specifically show the effects of path-loss, training overhead and network topology. 

\subsection{Baseline Schemes}\label{subsection:baseline}
In this section, we provide the effective average sum-rates of \ac{TDMA} and \ac{SU-MIMO}, which will be subsequenctly serve as a benchmark.
\begin{lemma}
The effective average sum-rate of \ac{TDMA} is:
\begin{equation}\label{eq:tdma}
\bar{R}_{\text{sum}}^{\textsc{tdma}} = \frac{1}{K}\frac{\tau_{\text{coh}} - KN_t}{\tau_{\text{coh}}} \ \sum_{k=1}^Kf\left(\textsf{snr}_k, N_t\right)
\end{equation}
where 
\\
\[
\textsf{snr}_k = \frac{P\eta_{kk}(1-\sigma_{\mathcal{E}_{kk}}^2)}{ P\eta_{kk}\sigma^2_{\mathcal{E}_{kk}} + \sigma_0^2} \]
\\
\end{lemma}

\begin{lemma}
The effective average sum-rate of \ac{SU-MIMO} is:
\begin{equation}\label{eq:sumimo}
\bar{R}_{\text{sum}}^{\textsc{su-mimo}} = \frac{\tau_{\text{coh}} - KN_t}{\tau_{\text{coh}}} \ \sum_{k=1}^Kf\left(\textsf{snr}_k, N_t\right)
\end{equation}

where
\[
\textsf{snr}_k = \frac{P\eta_{kk}(1-\sigma_{\mathcal{E}_{kk}}^2)}{ P\eta_{kk}\sigma_{\mathcal{E}_{kk}}^2 +\sum_{i\neq k}^K P\eta_{ki} + \sigma_0^2}
\]\\
\end{lemma}
The proof of both \eqref{eq:tdma} and \eqref{eq:sumimo} is a direct result of the mutual information bound derived in Appendix \ref{section:appendix}.

\subsection{Network Geometries} 
We are interested in assessing the performance of \ac{IA} in dense network scenarios. Therefore, we consider one transmitter/receiver pair per $25\ \text{m}^2$, as stated in \cite{80211ax}. More particularly, each node pair is placed inside a square of dimensions $5\ \text{m}\times 5\ \text{m}$, with transmitters being located at the center of their square and receivers being uniformly distributed at random within them. The $K$ squares are arranged in either a line or a rectangular grid. Finally, we also examine a complete random node deployment in a single square of the same size as the overall area of the grid arrangement, where there are more transmitters than receivers. In this case,  each receiver is associated to its closest transmitter and the rest of the transmitters remain idle. A realization of each topology is depicted in the figures below.

\begin{figure}[H]
\centering
\includegraphics[scale=0.7]{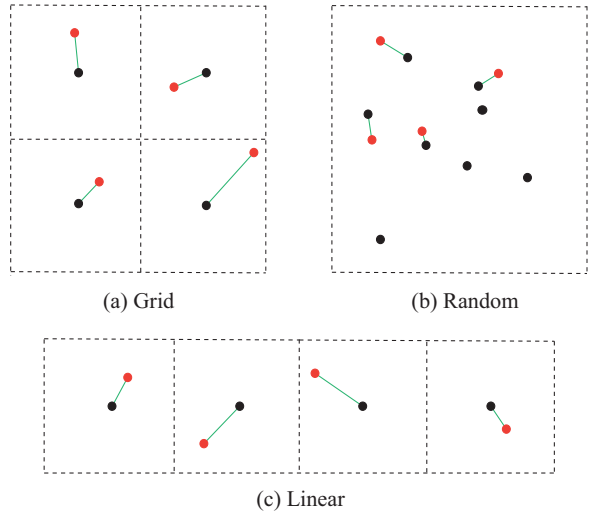}
\caption{Examples of network topologies. Red and black dots represent the locations of the receivers and transmitters, respectively.}
\end{figure}

\subsection{Large-Scale Fading Model}
The large-scale fading accounts for path-loss, which is modeled by the log-distance model
\[
\text{PL}_{ki} = 30 + 10\gamma\log_{10}(r_{ki}) = 10\log_{10}\frac{P_t}{P_{r,ki}} \quad (\text{dB})
\]
where $30\ \text{dB}$ is the loss at the reference distance of $1\text{m}$ and $\gamma$ is the path-loss exponent and $r_{ki}$ is the distance between the node pair $(k,i)$. The received power at each link $(k,i)$ is given by
\[
P_{r,ki} = W P \cdot  \eta_{ki} = P_t \cdot 10^{-\text{PL}_{ki}/10} \quad (\text{mW})
\]
where $W$ is the channel bandwidth.
\subsection{Numerical Results}
For given positions of the nodes, we compute the effective \ac{SNR} at receiver $k$ as
\\
\begin{align*}
\textsf{snr}_k &= \frac{P\eta_{kk}(1-\sigma_{\mathcal{E}_{kk}}^2)}{\sum_{i=1}^KP\eta_{ki}\sigma_{\mathcal{E}_{ki}}^2 + \sigma_0^2} \\
&=\frac{WP\eta_{kk}(1-\sigma_{\mathcal{E}_{kk}}^2)}{\sum_{i=1}^KWP\eta_{ki}\sigma_{\mathcal{E}_{ki}}^2 + W\sigma_0^2}\\
		&= \frac{P_{r,kk}(1-\sigma_{\mathcal{E}_{kk}}^2)}{\sum_{i=1}^KP_{r,ki}\sigma_{\mathcal{E}_{ki}}^2 + P_{\text{noise}}}
\end{align*}
\\
where the channel estimation error variance $\sigma_{\mathcal{E}_{ki}}^2$ with minimum training is given by
\\
\begin{align*}
\sigma_{\mathcal{E}_{ki}}^2 &= \frac{\sigma_0^2}{\sigma_0^2 +K P\eta_{ki}} \\
&= \frac{W\sigma_0^2}{W\sigma_0^2 +K W P\eta_{ki}}\\
	&= \frac{P_{\text{noise}}}{P_{\text{noise}} +K P_{r,ki}}.
\end{align*} 
\\
The noise power $P_{\text{noise}}$ is set to $-95\ \text{dBm}$. Then, using \eqref{eq:se_ia}, we calculate the spectral efficiency for one network realization. The expected value of the spectral efficiency is finally evaluated by considering $100$ realizations of the node locations and taking the average.

We further consider a symmetric network where all nodes are equipped with the same number of antennas, i.e., $N_r = N_t = N$. The value of $N$ is determined such that the \ac{IA} feasibility condition, $2N \geq d(K+1)$, is satisfied \cite{IA_KUser_IFC}.

\begin{figure}[H]
\centering
\includegraphics[scale = 0.6]{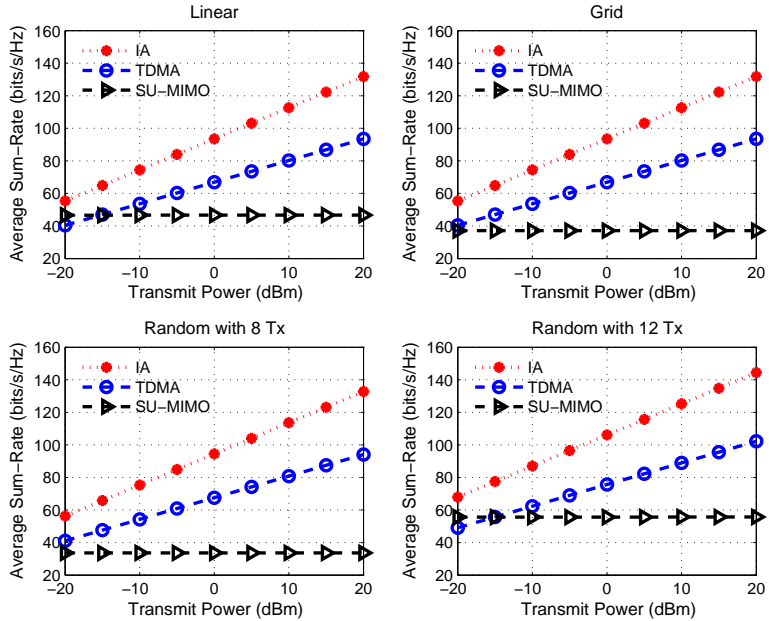}
\caption{Achievable sum-rate for the 4-user $5\times 5$ MIMO interference channel with imperfect \ac{CSI} and minimum pilot overhead. In \ac{IA}, each user attains 2 interference-free spatial streams. The overall network area is $100\ \text{m}^2$, $\gamma = 3.2$ and $\tau_{\text{coh}} = 100$ symbols.}
\label{fig:results1}
\end{figure}
In Fig. \ref{fig:results1}, we fixed the path-loss exponent, channel coherence time and network size to examine how the network geometry affects the spectral efficiency. As we notice, \ac{IA} performs similarly in all cases and attains higher performance than \ac{SU-MIMO} and \ac{TDMA} for all \ac{SNR} values. Furthermore, \ac{IA} achieves the best spectral efficiency in the case of a random topology with $12$ transmitters. This is because with multiple transmitters, each receiver is closer to its associated transmitter, yielding better \ac{SNR}.

\begin{figure}[H]
\centering
\includegraphics[scale = 0.6]{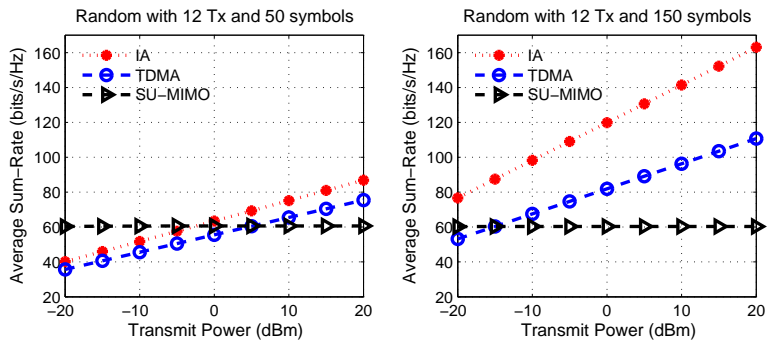}
\caption{Achievable sum-rate for the 4-user $5\times 5$ MIMO interference channel with imperfect \ac{CSI} and minimum pilot overhead. In \ac{IA}, each user attains 2 interference-free spatial streams. The overall network area is $100\ \text{m}^2$ and $\gamma = 3.2$.}
\label{fig:results2}
\end{figure}
We now choose the random topology with 12 transmitters to investigate the effect of training overhead and \ac{CSI} quality, i.e., variance of estimation error, on the performance of \ac{IA}. We first vary the channel coherence interval while keeping the path-loss exponent constant. From Fig. \ref{fig:results2}, we see that decreasing the coherence interval to $50$ symbols, the spectral efficiency of \ac{IA} deteriorates significantly due to the high \ac{CSI} acquisition overhead, i.e., half of the coherence interval is spent on training with $K(N_r + d) = 4(5+2) = 24$ pilot symbols. On the other hand, for sufficiently large coherence interval, i.e., $150$ symbols, \ac{IA} clearly outperforms both \ac{TDMA} and \ac{SU-MIMO}.

\begin{figure}[H]
\centering
\includegraphics[scale = 0.6]{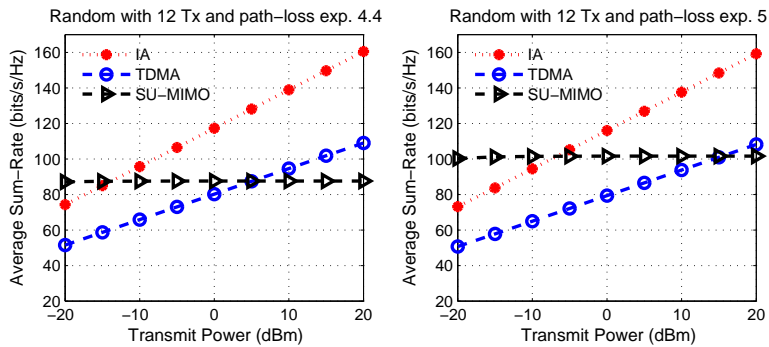}
\caption{Achievable sum-rate for the 4-user $5\times 5$ MIMO interference channel with imperfect \ac{CSI} and minimum pilot overhead. In \ac{IA}, each user attains 2 interference-free spatial streams. The overall network area is $100\ \text{m}^2$ and $\tau_{\text{coh}}= 150$ symbols.}
\label{fig:results3}
\end{figure}
Finally, we keep constant the coherence time and increase the path-loss exponent, which affects the \ac{CSI} quality. As we observe from Fig. \ref{fig:results3} the performance of \ac{IA} in dense scenarios is not considerably 
affected by the path-loss. 

\section{Conclusions}
In this paper, we obtained a closed-form expression for the average achievable sum-rate of \ac{MIMO} \ac{IA}, when communication channels are characterized by Rayleigh block-fading and path-loss, and \ac{CSI} is acquired through training. Our results show that \ac{IA} can boost the spectral efficiency in dense wireless networks under imperfect channel knowledge, and various levels of block fading and path-loss.

\appendices
\section{Lower Bound on the Mutual Information with Imperfect \ac{CSI}}\label{section:appendix}
Consider a \ac{SU-MIMO} system, where the transmitter and receiver are equipped with $N_t$ and $N_r$ antennas, respectively. The signal at the receiver, under channel estimation errors, is written as
\begin{align*}
\mathbf{y} = \sqrt{\frac{P}{N_t}}\mathbf{H}\mathbf{s} + \mathbf{n} = \sqrt{\frac{P}{N_t}}\hat{\mathbf{H}}\mathbf{s} + \tilde{\mathbf{n}}
\end{align*}
where $\tilde{\mathbf{n}} = \sqrt{\frac{P}{N_t}}\mathcal{E}\mathbf{s} + \mathbf{n}$ is the effective noise. We focus on the \ac{MI} $I(\mathbf{s};\mathbf{y} | \hat{\mathbf{H}})$, given the estimated channel $\hat{\mathbf{H}}$. Expanding conditional \ac{MI} into differential entropies yields

\begin{equation*}
I(\mathbf{s};\mathbf{y} | \hat{\mathbf{H}}) = h(\mathbf{s}| \hat{\mathbf{H}}) -  h(\mathbf{s}|\mathbf{y}, \hat{\mathbf{H}}) 
\end{equation*}

We choose $\mathbf{s}$ to be Gaussian given the estimate $\hat{\mathbf{H}}$ with covariance matrix $\mathbf{R}_{\mathbf{s}| \hat{\mathbf{H}}} =\mathbf{I}_{N_t}$. Thus, we have 
\[ h(\mathbf{s}| \hat{\mathbf{H}}) = \log\text{det}(\pi e\mathbf{R}_{\mathbf{s}| \hat{\mathbf{H}}}) = \log\text{det}(\pi e \mathbf{I}_{N_t}).
\]
We can derive a lower bound on \ac{MI} by upper-bounding $ h(\mathbf{s}|\mathbf{y}, \hat{\mathbf{H}})$ for any distribution on $\tilde{\mathbf{n}}$. We use the facts that 1) adding a constant does not change differential entropy, 2) conditioning always decreases entropy and 3) the entropy of a random variable with a given variance is upper-bounded by the entropy of a Gaussian random variable with the same variance (i.e., worst-case noise distribution). Hence,
\begin{align}\label{dif_entr_ineq}
h(\mathbf{s}|\mathbf{y}, \hat{\mathbf{H}})= h(\mathbf{s}-\mathbf{A}\mathbf{y}|\mathbf{y}, \hat{\mathbf{H}}) &\leq h(\mathbf{s}-\mathbf{A}\mathbf{y}|\hat{\mathbf{H}})\nonumber \\
&\leq \log\text{det}\left(\pi e \mathbf{R}_{\mathbf{s}-\mathbf{A}\mathbf{y}|\hat{\mathbf{H}}}\right)
\end{align}
Since inequality \eqref{dif_entr_ineq} holds for any given matrix, we pick $\mathbf{A}$ such that $\mathbf{A}\mathbf{y}$ is the \ac{MMSE} of $\mathbf{s}$ given $\hat{\mathbf{H}}$, in order to tighten the bound. Therefore, we have\footnote{We drop the subscript $\hat{\mathbf{H}}$ from the covariance matrices for ease of notation, e.g., $ \mathbf{R}_{\mathbf{sy}|\hat{\mathbf{H}}} = \mathbf{R}_{\mathbf{sy}}$.}
\begin{equation*}
\mathbf{A}\mathbf{y} = \mathbf{R}_{\mathbf{sy}} \mathbf{R}_{\mathbf{y}}^{-1}\mathbf{y}
\end{equation*}
and the estimation error is 
\begin{equation*}
\mathbf{s} - \mathbf{A}\mathbf{y} = \mathbf{s} -  \mathbf{R}_{\mathbf{sy}} \mathbf{R}_{\mathbf{y}}^{-1}\mathbf{y}
\end{equation*}
whereas its covariance matrix is given by
\begin{align}\label{est_error_covar1}
\mathbf{R}_{\mathbf{s}-\mathbf{A}\mathbf{y}|\hat{\mathbf{H}}} &= \mathbb{E}\left[( \mathbf{s} -  \mathbf{R}_{\mathbf{sy}} \mathbf{R}_{\mathbf{y}}^{-1}\mathbf{y})( \mathbf{s} -  \mathbf{R}_{\mathbf{sy}} \mathbf{R}_{\mathbf{y}}^{-1}\mathbf{y})^*| \hat{\mathbf{H}}\right]\nonumber \\
& = \mathbf{R}_{\mathbf{s}} - \mathbf{R}_{\mathbf{s}\mathbf{y}} \mathbf{R}_{\mathbf{y}}^{-1} \mathbf{R}_{\mathbf{y}\mathbf{s}} \nonumber\\
& = \mathbf{I}_{N_t} - \frac{P}{N_t}\hat{\mathbf{H}}^*\left(\frac{P}{N_t} \hat{\mathbf{H}}\hat{\mathbf{H}}^* + \mathbf{R}_{\tilde{\mathbf{n}}}\right)^{-1}\hat{\mathbf{H}} \nonumber
\end{align}
Using Woodbury matrix identity, the covariance matrix is given by
\begin{align*}
\mathbf{R}_{\mathbf{s}-\mathbf{A}\mathbf{y}|\hat{\mathbf{H}}} &= \mathbf{I}_{N_t} - \frac{P}{N_t}\hat{\mathbf{H}}^*\left(\frac{P}{N_t} \hat{\mathbf{H}}\hat{\mathbf{H}}^* + \mathbf{R}_{\tilde{\mathbf{n}}}\right)^{-1}\hat{\mathbf{H}}\\
& = \left(\mathbf{I}_{N_t} +\frac{P}{N_t} \hat{\mathbf{H}}^*\mathbf{R}_{\tilde{\mathbf{n}}}^{-1}\hat{\mathbf{H}}\right)^{-1}
\end{align*}

Finally, \eqref{dif_entr_ineq} is expressed as
\begin{equation*}
h(\mathbf{s}|\mathbf{y}, \hat{\mathbf{H}}) \leq \log\text{det}\ \pi e \left(\mathbf{I}_{N_t} + \frac{P}{N_t}\hat{\mathbf{H}}^*\mathbf{R}_{\tilde{\mathbf{n}}}^{-1}\hat{\mathbf{H}}\right)^{-1}
\end{equation*}
and the lower bound on conditional mutual information is
\begin{align*}\label{lb_mi_1}
I(\mathbf{s};\mathbf{y} | \hat{\mathbf{H}}) &\geq \log\text{det} \ \pi e \mathbf{I}_{N_t} - \log\text{det}\ \pi e \left(\mathbf{I}_{N_t} + \frac{P}{N_t}\hat{\mathbf{H}}^*\mathbf{R}_{\tilde{\mathbf{n}}}^{-1}\hat{\mathbf{H}}\right)^{-1} \nonumber \\
& = \log\text{det}\left(\mathbf{I}_{N_t} + \frac{P}{N_t}\hat{\mathbf{H}}^*\mathbf{R}_{\tilde{\mathbf{n}}}^{-1}\hat{\mathbf{H}}\right)\nonumber \\
& = \log\text{det}\left(\mathbf{I}_{N_r} +\frac{P}{N_t}\mathbf{R}_{\tilde{\mathbf{n}}}^{-1}\hat{\mathbf{H}} \hat{\mathbf{H}}^*\right)\\
& = \log\text{det}\left(\mathbf{I}_{N_r} +\frac{P(1-\sigma_{\mathcal{E}}^2)}{N_t(P\sigma^2_{\mathcal{E}} + \sigma_0^2)}\tilde{\mathbf{H}} \tilde{\mathbf{H}}^*\right)
\end{align*}
where $\tilde{\mathbf{H}} = \frac{1}{\sqrt{1-\sigma_{\mathcal{E}}^2}}\hat{\mathbf{H}}$ being the normalized estimated channel with unit variance, and $\mathbf{R}_{\tilde{\mathbf{n}}} = (P\sigma^2_{\mathcal{E}} + \sigma_0^2)\mathbf{I}_{N_r}$.


\end{document}